# The dynamics of thin gas layer moving between two fluids

Ivan V. Kazachkov[1,2]

[1]Dept of Information Technologies and Data Analysis,
Nizhyn Gogol State University, UKRAINE, http://www.ndu.edu.ua
[2]Dept of Energy Technology, Royal Institute of Technology, Stockholm, 10044, SWEDEN,
Ivan.Kazachkov@energy.kth.se, http://www.kth.se/itm/inst?l=en_UK

**Abstract**

The dynamics and stability of a thin gas layer moving between two fluid layers moving in the same or opposite direction is studied. The linear evolutionary equations describing the spatial-temporal dynamics of the interface's perturbations between gas and two fluid layers are derived for the flat two-dimensional case. Integral correlations across the layer are obtained, and the various kinds of time dependent base states are found. A linear stability is considered for the system using non-stationary equation array derived. The equation array consists of the two one-dimensional non-stationary equations of a seventh and fourth order. The results of the numerical study of the governing evolution equations support the results of the analysis for more simple limit cases. It is found that the thin sheet gas flow in-between two liquid layers is unstable and the peculiarities are found and discussed together with some applications available.

**Keywords:** gas thin layer, two liquid layers, interface, linear stability, three layers, deformable boundary.

## 1. Introduction

Thin gas or vapor sheets flows are often encountered in various experimental settings and technological applications: jet penetration into the pool of other liquid with a gas entrainment from the free surface [1-3] of a pool or from vaporization of a volatile coolant in a hot pool [4, 5]. Also, as a stage of a drop disintegration through the phase of a film flow [6], etc. In fact, the first two phenomena may occur together in a real case [4, 5]. Such kind liquid - gas (vapor) interfaces are prone to different types of instability being subjected to the influence of diverse physical factors and parameters, e.g. thicknesses of the layers, physical properties (viscosity, capillarity, etc.), velocities of the phases.

The jet's stability increases in infinite medium by increasing both viscosities of a jet and medium [7, 8]. And in case of a compound jet of the drip fluids, the surrounding fluid, having high enough viscosity can suppress the growing perturbations of the core so much that the instability of the internal jet and compound jet is fully predetermined by external jet parameters, its viscosity and thickness.

The instability of a thin gas (vapor) sheet between two fluid layers which may be, in general, countercurrent, was not reported in the literature yet. The aim of this paper is to clarify the phenomenon, to develop the mathematical model and analyze the physics of the task as much as possible. The linear evolution equations describing a spatial-temporal dynamics of the gas (vapor) - liquid interfaces are derived and the boundary conditions are stated. Then an appropriate integration of the equations is considered and the equations for the interface dynamics are obtained. The equation array is solved numerically and analytically (for same limit cases), and several sets of the base states are found, and their linear stability properties are examined.

## 2. Problem formulation

Two dimensional three-layer flow is considered for the physical situation shown in Fig. 1. In equilibrium state supposed to be three layers moving with constant velocities in different directions. The lowest layer is considered being in the rest or the coordinate system is touched with it moving with the same velocity. So that thin gas (vapor) layer is moving with respect to lower layer with velocity $U_1$. And the upper liquid layer moves with velocity $U_2$ against gas layer or in the same direction ($U_2 < 0$).

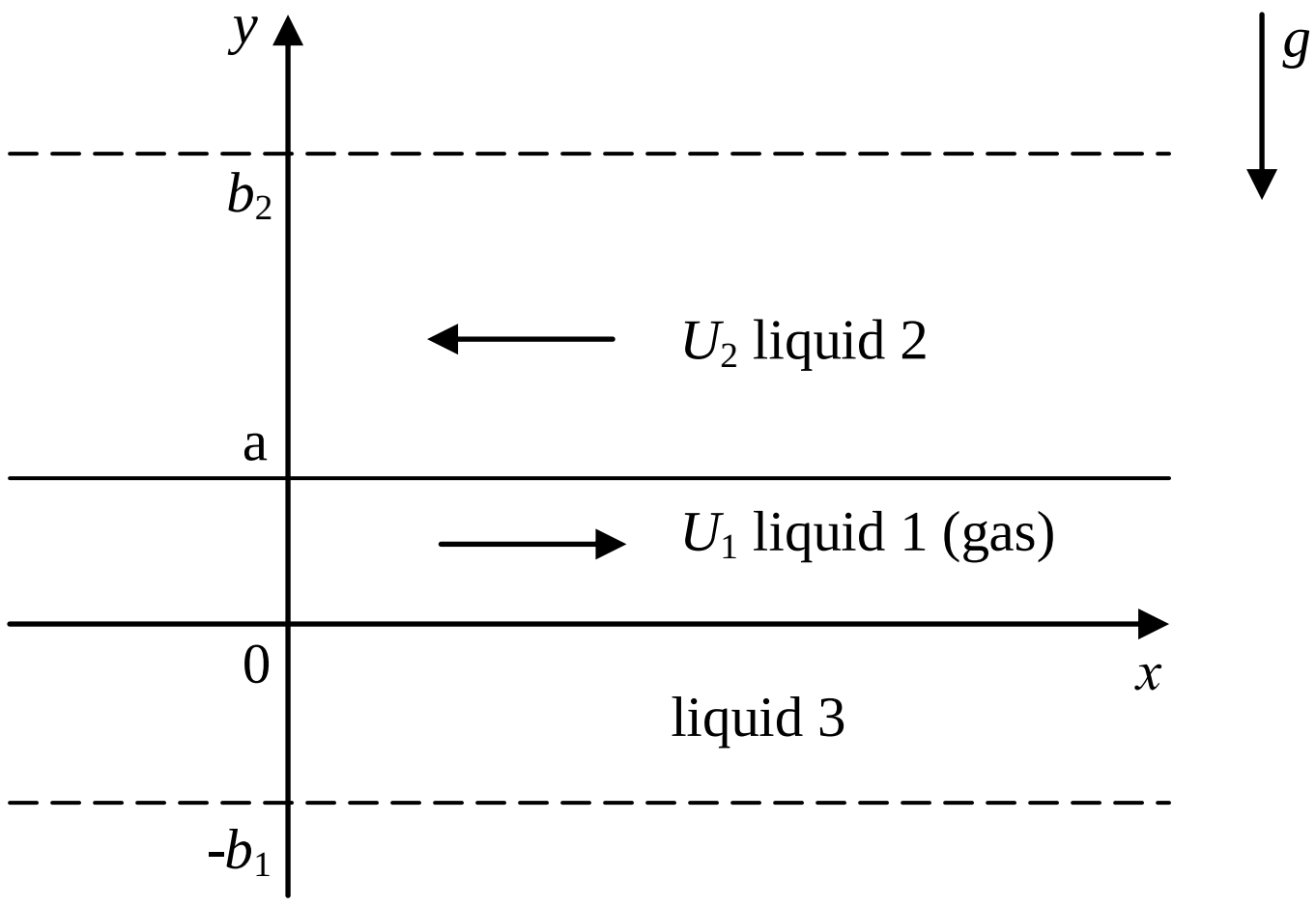

Fig. 1 The scheme of a three-layer flow: $y=0$, $y=a$ – free surfaces of a gas layer; $y=-b_1$, $y=b_2$ – free surface of the lower and top liquid layers, $U_1$, $U_2$ – flow velocities

It is assumed that inertia forces are big enough to neglect gravitational forces. And $y=0$, $y=a$ are the unperturbed interfaces of gas - liquid flows at the beginning. Gas (vapor) flow supposed to be incompressible.

The governing equations are the continuity and the momentum Navier-Stokes equations, which can be represented in the following linearized form:

$$\nabla\cdot\vec{V}_j=0,\qquad \rho_j(\partial_t\vec{V}_j+\vec{V}_j\nabla\vec{V}_j)=-\nabla p_j+\mu_j\nabla^2\vec{V}_j, \qquad (1)$$

where $\vec{V}=\{U, V\}$ is the fluid velocity field, $p$ is the pressure, $t$ is time, $\nabla\equiv(\partial_x,\partial_y)$, $\rho$ - density, $\mu$ - dynamic viscosity and indexes, $j$ = 1, 2, 3 are used for gas (vapor), top liquid and bottom liquid, respectively.

The boundary conditions are following: tangential stresses supposed to be negligible at the gas – liquid interfaces, therefore

$$y=h_1,\ \frac{\partial U_3}{\partial y}=-\frac{\partial V_3}{\partial x},\ \ \frac{\partial U_1}{\partial y}=-\frac{\partial V_1}{\partial x}; \qquad (2)$$

$$y=a+h_2,\ \frac{\partial U_2}{\partial y}=-\frac{\partial V_2}{\partial x},\ \ \frac{\partial U_1}{\partial y}=-\frac{\partial V_1}{\partial x}, \qquad (3)$$

where $h_j(x,t)$ are small-amplitude perturbations of the interfaces of the gas flow layer. The balances of the normal stresses at the interfaces and the augmented kinematics condition are:

$$y=h_1,\ v_1=v_3=\frac{\partial h_1}{\partial t}+U_1\frac{\partial h_1}{\partial x},\ p_1-p_3=\sigma_3\frac{\partial^2 h_1}{\partial x^2}+2(\mu_1\frac{\partial v_1}{\partial y}-\mu_3\frac{\partial v_3}{\partial y}); \qquad (4)$$

$$y=a+h_2,\ v_1=v_2=\frac{\partial h_2}{\partial t}+(U_1-U_2)\frac{\partial h_2}{\partial x},\ p_1-p_2=-\sigma_2\frac{\partial^2 h_2}{\partial x^2}+2(\mu_1\frac{\partial v_1}{\partial y}-\mu_2\frac{\partial v_2}{\partial y}), \qquad (5)$$

where $\sigma_2$, $\sigma_3$ are the surface tension coefficients for the low and top liquid layers with a gas layer, respectively. In the conditions (4) and (5) the capillary forces are taken with the opposite signs because convex and concave for the top and lower surfaces result in capillary pressure in a gas layer different by sign. No-slip is considered at the interfaces:

$$y = h_1,\ U_1 = U_3; \qquad y = a + h_2,\ U_1 = U_2; \tag{6), (7}$$

We suppose that in unperturbed state when the interfaces are the straightforward lines there is a gas slip at the interfaces. But no-slip is considered by gas flow with the perturbed interfaces having uneven boundaries. The liquid layers supposed to be thick enough to suppress the perturbation inside them:

$$y = -b_1,\ U_3 = V_3 = 0,\quad p_3 = 0; \tag{8}$$

$$y = b_2,\ U_2 = V_2 = 0,\quad p_2 = 0, \tag{9}$$

where are: $b_1 >> h_1, b_2 >> h_2$. The boundary conditions are linear in assumption that the long-wave small-amplitude perturbations of the interfaces are considered. And the boundary layer approximation may be applied for the thin gas layer dynamics. Then for a gas flow $p_1 = p_1(x,t)$, and the momentum equation in $y$ -direction is omitted.

Considering the instability of the interfaces one can integrate the equations (1) with boundary conditions (2) - (9) with respect to y and reduce the boundary problem (1) - (9) to the evolutionary equations for $h_j(x,t)$. For this purpose, further investigation is better to do in a dimensionless form. The scale values are chosen as follows : $a$, $U_1$, $a/U_1$, $\rho_1 U_1^2$ - for the length, velocity, time and pressure, respectively. It is considered for simplicity that in the unperturbed state the layers move with the constant their velocities along the axis $x$. Then the dimensionless boundary problem (1) – (9) for perturbations is got in the following form:

$$\frac{\partial u_2}{\partial t} - U_{21}\frac{\partial u_2}{\partial x} = -\rho_{12}\frac{\partial p_2}{\partial x} + \frac{\nu_{21}}{\mathrm{Re}}\left(\frac{\partial^2 u_2}{\partial x^2} + \frac{\partial^2 u_2}{\partial y^2}\right),\quad \frac{\partial v_2}{\partial t} - U_{21}\frac{\partial v_2}{\partial x} = -\rho_{12}\frac{\partial p_2}{\partial y} + \frac{\nu_{21}}{\mathrm{Re}}\left(\frac{\partial^2 v_2}{\partial x} + \frac{\partial^2 v_2}{\partial y}\right),$$

$$\frac{\partial u_j}{\partial x} = -\frac{\partial v_j}{\partial y},\quad \frac{\partial u_1}{\partial t} + \frac{\partial u_1}{\partial x} = -\frac{\partial p_1}{\partial x} + \frac{1}{\mathrm{Re}}\left(\frac{\partial^2 u_1}{\partial x^2} + \frac{\partial^2 u_1}{\partial y^2}\right), \tag{10}$$

$$\frac{\partial u_3}{\partial t} = -\rho_{13}\frac{\partial p_3}{\partial x} + \frac{\nu_{31}}{\mathrm{Re}}\left(\frac{\partial^2 u_3}{\partial x^2} + \frac{\partial^2 u_3}{\partial y^2}\right),\quad \frac{\partial v_3}{\partial t} = -\rho_{13}\frac{\partial p_3}{\partial y} + \frac{\nu_{31}}{\mathrm{Re}}\left(\frac{\partial^2 v_3}{\partial x^2} + \frac{\partial^2 v_3}{\partial y^2}\right),$$

where the momentum equation for the gas flow is omitted because a boundary layer approach is adopted for it due to considered thin gas layer. Here $\mathrm{Re} = U_1 a/\nu_1$ - the Reynolds number for a gas flow, $\nu$ - kinematic viscosity coefficient, $\rho_{12} = \rho_1/\rho_2$, $\rho_{13} = \rho_1/\rho_3$, $\nu_{21} = \nu_2/\nu_1$, $\nu_{31} = \nu_3/\nu_1$, $U_{21} = U_2/U_1$. Here $U_{21}$ characterizes the liquid to gas velocity ratio, which supposed to play an important role in an interfacial instability. The dimensionless boundary conditions (2) – (9) are going to the following:

$$y = \varsigma_1,\ \frac{\partial u_3}{\partial y} = -\frac{\partial v_3}{\partial x},\ \frac{\partial u_1}{\partial y} = -\frac{\partial v_1}{\partial x}; \qquad y = 1 + \varsigma_2,\ \frac{\partial u_2}{\partial y} = -\frac{\partial v_2}{\partial x},\ \frac{\partial u_1}{\partial y} = -\frac{\partial v_1}{\partial x}; \tag{11), (12}$$

$$y=\varsigma_1,\quad v_1=v_3=\frac{\partial\varsigma_1}{\partial t}+\frac{\partial\varsigma_1}{\partial x},\quad p_1-\rho_{31}p_3=\frac{1}{We_3}\frac{\partial^2\varsigma_1}{\partial x^2}+\frac{2}{\mathrm{Re}}\left(\frac{\partial v_1}{\partial y}-\mu_{31}\frac{\partial v_3}{\partial y}\right); \tag{13}$$

$$y=1+\varsigma_2,\quad v_1=v_2=\frac{\partial\varsigma_2}{\partial t}+(1-v_{21})\frac{\partial\varsigma_2}{\partial x},\quad p_1-\rho_{21}p_2=-\frac{1}{We_2}\frac{\partial^2\varsigma_2}{\partial x^2}+\frac{2}{\mathrm{Re}_1}\left(\frac{\partial v_1}{\partial y}-\mu_{21}\frac{\partial v_2}{\partial y}\right); \tag{14}$$

$$y=\varsigma_1,\quad u_1=u_3;\qquad y=1+\varsigma_2,\quad u_1=u_2. \tag{15}$$

$$y=\beta_2,\quad u_2=v_2=p_2=0;\qquad y=-\beta_1,\quad u_3=v_3=p_3=0; \tag{16), (17}$$

Here are: $\varsigma_j=h_j/a$, $\varsigma_2=h_2/a$, $We_3=\rho_1U_1^2a/\sigma_3$, $We_2=\rho_1U_1^2a/\sigma_2$ - the Weber's numbers for a gas flow boundary with the top and bottom liquid layers, respectively, $\sigma_2,\sigma_3$- the surface tension coefficients, $We_3=We_2\sigma_{23}$, $\sigma_{23}=\sigma_2/\sigma_3$, $\mu_{21}=\mu_2/\mu_1$, $\mu_{31}=\mu_3/\mu_1$, $\beta_1=b_1/a$, $\beta_2=b_2/a$, $\beta_j >> \zeta_j$.

## 3. Derivation of the evolutionary equations for oscillations of the gas layer boundaries

Considering the stability problem for the two liquid-gas interfaces one can get the evolutionary equations based on the dimensionless boundary problem (10) – (17) stated above. With integration of all equations (10) across the layers with respect to *y* in corresponding ranges gives:

$$\int_{\varsigma_1}^{1+\varsigma_2}\frac{\partial u_1}{\partial t}dy+\int_{\varsigma_1}^{1+\varsigma_2}\frac{\partial u_1}{\partial x}dy=-\int_{\varsigma_1}^{1+\varsigma_2}\frac{\partial p_1}{\partial x}dy+\frac{1}{\mathrm{Re}_1}\int_{\varsigma_1}^{1+\varsigma_2}\left(\frac{\partial^2u_1}{\partial x^2}+\frac{\partial^2u_1}{\partial y^2}\right)dy,$$

$$\int_{1+\varsigma_2}^{\beta_2}\frac{\partial u_2}{\partial x}dy=-\int_{1+\varsigma_2}^{\beta_2}\frac{\partial v_2}{\partial y}dy,\quad \int_{\varsigma_1}^{1+\varsigma_2}\frac{\partial u_1}{\partial x}dy=-\int_{\varsigma_1}^{1+\varsigma_2}\frac{\partial v_1}{\partial y}dy,\qquad p_1=p_1(x,t); \tag{18}$$

$$\int_{1+\varsigma_2}^{\beta_2}\frac{\partial u_2}{\partial t}dy-v_{21}\int_{1+\varsigma_2}^{\beta_2}\frac{\partial u_2}{\partial x}dy=-\int_{1+\varsigma_2}^{\beta_2}\frac{\partial p_2}{\partial x}dy+\frac{\rho_{21}}{\mathrm{Re}}\int_{1+\varsigma_2}^{\beta_2}\left(\frac{\partial^2u_2}{\partial x^2}+\frac{\partial^2u_2}{\partial y^2}\right)dy; \tag{19}$$

$$\int_{-\beta_1}^{\varsigma_1}\frac{\partial u_3}{\partial x}dy=-\int_{-\beta_1}^{\varsigma_1}\frac{\partial v_3}{\partial y}dy,\quad \int_{-\beta_1}^{\varsigma_1}\frac{\partial u_3}{\partial t}dy=-\int_{-\beta_1}^{\varsigma_1}\frac{\partial p_3}{\partial x}dy+\frac{\rho_{31}}{\mathrm{Re}_3}\int_{-\beta_1}^{\varsigma_1}\left(\frac{\partial^2u_3}{\partial x^2}+\frac{\partial^2u_3}{\partial y^2}\right)dy. \tag{20}$$

The integral correlations (18) – (20) are written for a gas layer, and for the top and bottom liquid layers, respectively. The momentum conservation equations for the transversal components were not integrated yet. They are to be used further in a differential form.

Then using the following transformation for the integrals with variable limits:

$$\int_{\alpha(g)}^{\beta(g)}\frac{\partial f}{\partial g}dy=\frac{\partial}{\partial g}\int_{\alpha(g)}^{\beta(g)}fdy+f(\alpha(g))\frac{\partial\alpha}{\partial g}-f(\beta(g))\frac{\partial\beta}{\partial g},$$

from (18) - (20) yields

$$\frac{\partial q_1}{\partial t}+(u_1)_{\varsigma_1}\frac{\partial \varsigma_1}{\partial t}-(u_1)_{1+\varsigma_1}\frac{\partial \varsigma_2}{\partial t}+\frac{\partial q_1}{\partial x}+(u_1)_{\varsigma_1}\frac{\partial \varsigma_1}{\partial x}-(u_1)_{1+\varsigma_2}\frac{\partial \varsigma_2}{\partial x}= \tag{21}$$

$$=\frac{\partial p_1}{\partial x}(\varsigma_1-\varsigma_2-1)+\frac{1}{\mathrm{Re}}\left[2\left(\frac{\partial v_1}{\partial x}\right)_{1+\varsigma_2}^{\varsigma_1}-\left(\frac{\partial v_1}{\partial y}\frac{\partial \varsigma_1}{\partial x}\right)_{1+\varsigma_2}^{\varsigma_1}\right],\ \frac{\partial q_1}{\partial x}=(v_1)_{1+\varsigma_2}^{\varsigma_1}+(u_1)_{1+\varsigma_2}\frac{\partial \varsigma_2}{\partial x}-(u_1)_{\varsigma_1}\frac{\partial \varsigma_1}{\partial x};$$

$$\frac{\partial q_2}{\partial x}=(v_2)_{1+\varsigma_2}-(u_2)_{1+\varsigma_2}\frac{\partial \varsigma_2}{\partial x},\qquad \frac{\partial q_2}{\partial t}-v_{21}\left[\frac{\partial q_2}{\partial x}+(u_2)_{1+\varsigma_2}\frac{\partial \varsigma_2}{\partial x}\right]+(u_2)_{1+\varsigma_2}\frac{\partial \varsigma_2}{\partial t}=$$

$$=-\int_{1+\varsigma_2}^{\beta_2}\frac{\partial p_2}{\partial x}dy+\frac{v_{21}}{\mathrm{Re}}\left[2\left(\frac{\partial v_2}{\partial x}\right)_{1+\varsigma_2}-\left(\frac{\partial v_2}{\partial y}\right)_{1+\varsigma_2}\frac{\partial \varsigma_2}{\partial x}\right]; \tag{22}$$

$$\frac{\partial q_3}{\partial x}+(v_3)_{\varsigma_1}=(u_3)_{\varsigma_1}\frac{\partial \varsigma_1}{\partial x},\ \frac{\partial q_3}{\partial t}-(u_3)_{\varsigma_1}\frac{\partial \varsigma_1}{\partial t}=-\int_{-\beta_1}^{\varsigma_1}\frac{\partial p_3}{\partial x}dy+\frac{v_{31}}{\mathrm{Re}}\left[2\left(\frac{\partial v_3}{\partial x}\right)_{\varsigma_1}-\left(\frac{\partial v_3}{\partial y}\right)_{\varsigma_1}\frac{\partial \varsigma_1}{\partial x}\right]. \tag{23}$$

Here are: $q_1=\int_{\varsigma_1}^{1+\varsigma_2}u_1dy$, $q_2=\int_{1+\varsigma_2}^{\beta_2}u_2dy$, $q_3=\int_{-\beta_1}^{\varsigma_1}u_3dy$ - the flow rates by gas and by two fluid layers, respectively. Then the main problem is to calculate the integral with pressure in (22), (23), and to calculate the normal stresses in the boundary conditions (13), (14). In a linear approach the terms of second order by perturbations should be omitted in (21) – (23), as well as the boundary values have to be substituted from the boundary conditions (11) – (17). The most difficult is to calculate the integral with pressure in (22) – (23) and to calculate the normal stresses in the boundary conditions.

To close the equation array (21) - (23) thus obtained one needs to know $\partial v/\partial y$ at the boundaries and the integrals of pressure in the two liquid layers. From the mass conservation equation yields $\partial v/\partial y=-\partial u/\partial x$, therefore transversal velocity is expressed as $v=-\int(\partial u/\partial x)dy$. And in the equations (21) - (23) the following expressions are got:

$$(v_1)_{\varsigma_1}^{1+\varsigma_2}=-\frac{\partial q_1}{\partial x},\ (v_2)_{1+\varsigma_2}=\frac{\partial q_2}{\partial x},\ (v_3)_{\varsigma_1}=-\frac{\partial q_3}{\partial x}.$$

And further $\partial v/\partial y$ should be expressed thought the functions $h_j(x,\ t)$ and $q_j(x,\ t)$, which are to be calculated later. For this, from the mass conservation equation and boundary conditions (11), (12) follows at the interfaces:

$$\frac{\partial^2 v}{\partial y^2}=-\frac{\partial}{\partial y}\left(\frac{\partial u}{\partial x}\right)=-\frac{\partial}{\partial x}\left(\frac{\partial u}{\partial y}\right)=-\frac{\partial}{\partial x}\left(-\frac{\partial v}{\partial x}\right)=\frac{\partial^2 v}{\partial x^2}, \tag{24}$$

and the same for any odd order derivation: $\partial^4 v/\partial y^4=\partial^4 v/\partial x^4$, etc. These correlations are satisfied only at the interfaces. Using (24) and the same for the fourth order correlations one can get the approximations of the fourth order for the transversal velocity components across the liquid layers. Let us consider, which is the most appropriate by these conditions.

**Polynomial approximations for transversal velocities' components**

Starting from the fourth-order approximation $v=c_1+c_2y+c_3y^2+c_4y^3$, using the 2 boundary conditions for $v$ and $\partial^2 v/\partial x^2$. Here $c_j$ – constants to be computed from the boundary conditions:
$y=\gamma_i$, $v=(v)_i$, $\left(\partial^2 v/\partial y^2\right)_i = \left(\partial^2 v/\partial x^2\right)_i$, where $i$=1,2, $\gamma_1$, $\gamma_2$ – are the bottom and top boundaries, e.g. for $v_1$: $\gamma_1=\zeta_1$, $\gamma_2=1+\zeta_2$, $(v)_1=v_1$ at $y=\zeta_1$, $(v_1)_1^2=(v_1)_{1+\varsigma_2}-(v_1)_{\varsigma_1}$, etc. Then for $i$=1,2: $c_1+c_2\gamma_i+c_3\gamma_i^2+c_4\gamma_i^3=(v)_i$, $c_3+3c_4\gamma_i=(\partial^2 v/\partial x^2)_i$. Thus, for each of 3 layers the system of 4 equations must be solved with specific values $\gamma_1$, $\gamma_2$, $(v)_{1,2}$ and $(\partial^2 v/\partial x^2)_{1,2}$ of the layers. In general, solution is

$$c_1=(v_1)+\frac{1}{2}\left(\frac{\partial^2 v}{\partial x^2}\right)_1\gamma_1\gamma_2-\frac{\gamma_1(v)_2^1}{\gamma_1-\gamma_2}+\frac{1}{2}\left(\frac{\partial^2 v}{\partial x^2}\right)_2^1\frac{\gamma_1\gamma_2(\gamma_2-2\gamma_1)}{3(\gamma_1-\gamma_2)},\quad c_3=\frac{1}{2}\left(\frac{\partial^2 v}{\partial x^2}\right)_1-\frac{\gamma_1}{2(\gamma_1-\gamma_2)}\left(\frac{\partial^2 v}{\partial x^2}\right)_2^1,$$

$$c_2=\frac{(v)_2^1}{\gamma_1-\gamma_2}-\frac{1}{2}\left(\frac{\partial^2 v}{\partial x^2}\right)_1(\gamma_1+\gamma_2)+\frac{1}{2}\left(\frac{\partial^2 v}{\partial x^2}\right)_2^1\frac{\gamma_2^3-3\gamma_1\gamma_2^2+2\gamma_1^3}{3(\gamma_1-\gamma_2)^2},\quad c_4=\frac{1}{6(\gamma_1-\gamma_2)}\left(\frac{\partial^2 v}{\partial x^2}\right)_2^1.$$

After substitution of specific values of $v$ and $\partial^2 v/\partial x^2$ from corresponding boundaries of the layers, yields in a linear approach:

$$c_{11}=(v_1)_{\varsigma_1},\quad c_{12}=(v_1)_{\varsigma_1}^{1+\varsigma_2}+\frac{1}{6}\left(\frac{\partial^2 v_1}{\partial x^2}\right)_{\varsigma_1}^{1+\varsigma_2}-\frac{1}{2}\left(\frac{\partial^2 v_1}{\partial x^2}\right)_{\varsigma_1},\quad c_{13}=\frac{1}{2}\left(\frac{\partial^2 v_1}{\partial x^2}\right)_{\varsigma_1},\quad c_{14}=-\frac{1}{6}\left(\frac{\partial^2 v_1}{\partial x^2}\right)_{\varsigma_1}^{1+\varsigma_2};$$

$$c_{21}=(v_2)_{1+\varsigma_2}+\frac{\beta_2}{2}\left(\frac{\partial^2 v_2}{\partial x^2}\right)_1+\frac{(v_2)_{1+\varsigma_2}}{\beta_2-1}+\frac{1}{6}\left(\frac{\partial^2 v_2}{\partial x^2}\right)_1\frac{\beta_2(\beta_2-2)}{(1-\beta_2)}=\frac{\beta_2}{\beta_2-1}(v_2)_{1+\varsigma_2}+\frac{\beta_2(1-2\beta_2)}{6(1-2\beta_2)}\left(\frac{\partial^2 v_2}{\partial x^2}\right)_{1+\varsigma_2},$$

$$c_{22}=\frac{(v_2)_{1+\varsigma_2}}{1-\beta_2}-\frac{2\beta_2^2+2\beta_2-1}{6(\beta_2-1)}\left(\frac{\partial^2 v_2}{\partial x^2}\right)_{1+\varsigma_2},\quad c_{23}=\frac{\beta_2}{2(\beta_2-1)}\left(\frac{\partial^2 v_2}{\partial x^2}\right)_{1+\varsigma_2},\quad c_{24}=\frac{1}{6(1-\beta_2)}\left(\frac{\partial^2 v_2}{\partial x^2}\right)_{1+\varsigma_2}; \qquad (25)$$

$$c_{31}=(v_3)_{\varsigma_1},\quad c_{32}=\frac{(v_3)_{\varsigma_1}}{\beta_1}+\frac{\beta_1}{3}\left(\frac{\partial^2 v_3}{\partial x^2}\right)_{\varsigma_1},\quad c_{33}=\frac{1}{2}\left(\frac{\partial^2 v_3}{\partial x^2}\right)_{\varsigma_1},\quad c_{34}=\frac{1}{6\beta_1}\left(\frac{\partial^2 v_3}{\partial x^2}\right)_{\varsigma_1}.$$

These approximations work well for thin liquid layers. If $\beta_i >> 1$, they do not fit well to our task, especially in case of at least one infinite liquid layer ($\beta_i = \infty$). Therefore, we will not use them further.

Because the interfacial perturbations supposed to die in the liquid layers as far as they go inside the layers from the interfaces. Therefore, it seems to be reasonable to compute the transversal velocity approximations in the following form:

$$v = d_1+\frac{d_2}{y}+\frac{d_3}{y^2}+\frac{d_4}{y^3}, \qquad (26)$$

where $d_i$ - const. The approximations (26) are written in a generalized form for both the top and bottom liquid layers with their own coefficients $d_i$. From (26) follows that in case of the infinite liquid layers there are to be determined only two constants because the other two are satisfied at the infinity automatically. Thus, by $\beta_i=\infty$ there are $d_1$=0 and

$$v=\frac{\gamma_1}{2y}\left[3(v)_1-\frac{1}{2}\left(\frac{\partial^2 v}{\partial x^2}\right)_1\gamma_1^2\right]+\frac{\gamma_1^2}{4y^2}\left[\left(\frac{\partial^2 v}{\partial x^2}\right)_1\gamma_1^2-2(v)_1\right], \qquad (27)$$

where are: $y=\gamma_1$- the interface between gas and liquid and $y=\gamma_2$- the faraway boundary of the liquid layer. Here and further the symbolic calculations are used for both the top and bottom layers simultaneously for the simplification. The boundary conditions like (24) are expressed in a symbolic form as

$$y=\gamma_i,\ \ v=(v)_i,\ \ \left(\partial^2 v/\partial y^2\right)_i=\left(\partial^2 v/\partial x^2\right)_i \tag{28}$$

where $i$ = 1,2 and $\gamma_1$, $\gamma_2$ are the bottom and the top boundaries, respectively, e.g. for $v_1$ should be $\gamma_1=\zeta_1$, $\gamma_2=1+\zeta_2$, $(v)_1=v_1$ by $y=\zeta_1$, $(v_1)_1^{\ 2}=(v_1)_{1+\zeta_2}-(v_1)_{\zeta_1}$, and so on.

Substitution of (26) into (28) with account of above mentioned yields

$$\begin{aligned}
d_1&=\frac{\gamma_1^2(4\gamma_2-\gamma_1)(v)_1}{(\gamma_2-\gamma_1)(\gamma_1^2-3\gamma_1\gamma_2+\gamma_2^2)}+\frac{\gamma_1^4(3\gamma_1-2\gamma_2)(v'')_1}{6(\gamma_2-\gamma_1)(\gamma_1^2-3\gamma_1\gamma_2+\gamma_2^2)},\\
d_2&=\frac{6\gamma_1^2\gamma_2^2(v)_1}{(\gamma_2-\gamma_1)(\gamma_1^2-3\gamma_1\gamma_2+\gamma_2^2)}+\frac{\gamma_1^4(\gamma_1^2+\gamma_1\gamma_2-\gamma_2^2)(v'')_1}{6(\gamma_2-\gamma_1)(\gamma_1^2-3\gamma_1\gamma_2+\gamma_2^2)},\\
d_3&=\frac{2\gamma_1^2\gamma_2^2(\gamma_1+\gamma_2)(v)_1}{(\gamma_2-\gamma_1)(\gamma_1^2-3\gamma_1\gamma_2+\gamma_2^2)}+\frac{\gamma_1^4\gamma_2(\gamma_1^2+\gamma_1\gamma_2-5\gamma_1^2)(v'')_1}{6(\gamma_2-\gamma_1)(\gamma_1^2-3\gamma_1\gamma_2+\gamma_2^2)},\\
d_4&=\frac{2\gamma_1^3\gamma_2^3(v)_1}{(\gamma_2-\gamma_1)(\gamma_1^2-3\gamma_1\gamma_2+\gamma_2^2)}+\frac{\gamma_1^5\gamma_2^2(\gamma_2-2\gamma_1)(v'')_1}{6(\gamma_2-\gamma_1)(\gamma_1^2-3\gamma_1\gamma_2+\gamma_2^2)}.
\end{aligned}\tag{29}$$

The specific values $d_i$ are obtained from (29) accounting that for lower liquid $\gamma_1=\varsigma_1$, $\gamma_2=-\beta_2$, and for the top $\gamma_1=1+\varsigma_2$, $\gamma_2=\beta_2$, respectively. After linearization yield the following. At the top ($y\in[1+\zeta_2, \beta_2]$):

$$\begin{aligned}
d_1&=\frac{(4\beta_2-1)(v_2)_{1+\varsigma_2}}{(\beta_2-1)(\beta_2^2-3\beta_2+1)}+\frac{(3-2\beta_2)(\partial^2 v/\partial x^2)_{1+\varsigma_2}}{6(\beta_2-1)(\beta_2^2-3\beta_2+1)},\\
d_2&=\frac{6\beta_2(v_2)_{1+\varsigma_2}}{(\beta_2-1)(\beta_2^2-3\beta_2+1)}+\frac{(\beta_2^2-\beta_2-1)(\partial^2 v/\partial x^2)_{1+\varsigma_2}}{2\beta_2(\beta_2^2-3\beta_2+1)},\\
d_3&=\frac{2\beta_2^2(\beta_2+1)(v_2)_{1+\varsigma_2}}{(\beta_2-1)(\beta_2^2-3\beta_2+1)}+\frac{\beta_2(\beta_2^2+\beta_2-5)(\partial^2 v/\partial x^2)_{1+\varsigma_2}}{6(1-\beta_2)(\beta_2^2-3\beta_2+1)},\\
d_4&=\frac{\beta_2^3(v_2)_{1+\varsigma_2}}{(1-\beta_2)(\beta_2^2-3\beta_2+1)}+\frac{\beta_2(\beta_2-2)(\partial^2 v/\partial x^2)_{1+\varsigma_2}}{6(\beta_2-1)(\beta_2^2-3\beta_2+1)}.
\end{aligned}\tag{30}$$

And by $\beta_2$ >> 1 the expressions (30) are simplified:

$$\begin{aligned}
&d_2=\frac{1}{\beta_2}\left[\frac{1}{2}\left(\frac{\partial^2 v_2}{\partial x^2}\right)_{1+\varsigma_2}-6(v_2)_{1+\varsigma_2}\right],\quad d_1=\frac{1}{\beta_2^2}\left[4(v_2)_{1+\varsigma_2}-\frac{1}{3}\left(\frac{\partial^2 v_2}{\partial x^2}\right)_{1+\varsigma_2}\right],\\
&d_3=2(v_2)_{1+\varsigma_2}-\frac{1}{6}\left(\frac{\partial^2 v_2}{\partial x^2}\right)_{1+\varsigma_2},\quad d_4=\frac{1}{6}\left(\frac{\partial^2 v_2}{\partial x^2}\right)_{1+\varsigma_2}-(v_2)_{1+\varsigma_2}.
\end{aligned}\tag{31}$$

In the lower part (lower liquid layer, $y\in[-\beta_1,\zeta_1]$):

$$d_4 = \varsigma_1^3\left[\frac{\varsigma_1^2}{6}\left(\frac{\partial^2 v_3}{\partial x^2}\right)_{\varsigma_1} - \left(1-\frac{4}{\beta_1}\varsigma_1+\frac{12}{\beta_1^2}\varsigma_1^2\right)(v_3)_{\varsigma_1}\right],\quad d_1 = \frac{4\varsigma_1^2}{\beta_1^2}(v_3)_{\varsigma_1} \approx 0$$ - the third-order by perturbations;

$$d_2 = 6(v_3)_{\varsigma_1}\frac{\varsigma_1^2}{\beta_1^2}(\beta_1 - 4\varsigma_1),\quad d_3 = \varsigma_1^2\left[2\left(1-\frac{5}{\beta_1}\varsigma_1+\frac{16\varsigma_1^2}{\beta_1^2}\right)(v_3)_{\varsigma_1} - \frac{\varsigma_1^2}{6}\left(\frac{\partial^2 v_3}{\partial x^2}\right)_{\varsigma_1}\right], \tag{32}$$

by $\beta_1 >> 1$ (thick layer), the expressions (32) are simplified:

$$d_1 \approx 0,\quad d_2 = \frac{6}{\beta_1}\varsigma_1^2(v_3)_{\varsigma_1},\quad d_3 = \varsigma_1^2\left[2(v_3)_{\varsigma_1} - \frac{\varsigma_1^2}{6}\left(\frac{\partial^2 v_3}{\partial x^2}\right)_{\varsigma_1}\right],\quad d_4 = \varsigma_1^3\left[\frac{\varsigma_1^2}{6}\left(\frac{\partial^2 v_3}{\partial x^2}\right)_{\varsigma_1} - (v_3)_{\varsigma_1}\right]. \tag{33}$$

Here $d_1 \approx 0$, though it is of the third order like $d_2$. But $d_2$ is divided by $\zeta_1$, while in a derivative $\partial^2 v_3 / \partial x^2$ - by $\varsigma_1^3$, therefore, this term is substantial but $d_1$ is a third-order constant (neglected).

To compute the integrals from pressure in liquid media, in the equations (22), (23), the momentum equations must be integrated in a transversal direction. Thus, integrating the (10) by $y$ in the range from $y$ to $\beta_i$, accounting the conditions of a suppression of all perturbations by $y=\beta_i$ (zero perturbations at $y=\beta_I$, they die in the thick layer). Therefore, for the profiles (26) yields:

$$p_i(y) = \frac{v_{i1}}{\mathrm{Re}}\left[\frac{\partial^2 c_1}{\partial x^2}y + \frac{\partial^2 c_2}{\partial x^2}\frac{y^2}{2} + \frac{\partial^2 c_3}{\partial x^2}\frac{y^3}{3} + \frac{\partial^2 c_4}{\partial x^2}\frac{y^4}{4} + 2\left(c_3 y + \frac{3}{2}c_4 y^2\right)\right] + \tag{34}$$
$$+(3-i)v_{21}\left(\frac{\partial c_1}{\partial x}y + \frac{\partial c_2}{\partial x}\frac{y^2}{2} + \frac{\partial c_3}{\partial x}\frac{y^3}{3} + \frac{\partial c_4}{\partial x}\frac{y^4}{4}\right) + \left(\frac{\partial c_1}{\partial t}y + \frac{\partial c_2}{\partial t}\frac{y^2}{2} + \frac{\partial c_3}{\partial t}\frac{y^3}{3} + \frac{\partial c_4}{\partial t}\frac{y^4}{4}\right) + p_i,$$

where $i$=2,3 for the top and lower liquid layers, correspondingly. In the equations (22), (23), the integrals $\int_{\gamma_1}^{\gamma_2}\left(\partial p_i / \partial x\right)dy$ are transformed with $\gamma_1=\zeta_1$ or $1+\zeta_2$, $\gamma_2=\beta_i$. And $p_i$ in (34) does not depend on $y$, therefore:

$$p_i = \frac{\gamma_2 v_{i1}}{2\,\mathrm{Re}}\left[\frac{\gamma_2+1}{\gamma_1-\gamma_2}\left(\frac{\partial^2 v}{\partial x^2}\right)_{\gamma_1} + \frac{\gamma_2-\gamma_1}{6}\left(\frac{\partial^4 v}{\partial x^4}\right)_{\gamma_1}\right] + \frac{\gamma_2}{2}\left(\frac{1}{\gamma_2-\gamma_1} + \frac{\gamma_2-\gamma_1}{6}\frac{\partial^2}{\partial x^2}\right)\left(\frac{\partial p_i}{\partial y}\right)_{\gamma_1}, \tag{35}$$

or, without using the pressure gradient on the boundary (can be expressed through $q_i(x)$ as done below):

$$p_i = \frac{v_{i1}}{\mathrm{Re}}\left[\frac{\gamma_2(\gamma_2-1)}{2(\gamma_1-\gamma_2)}\left(\frac{\partial^2 v}{\partial x^2}\right)_{\gamma_1} + \frac{\gamma_2}{12}(\gamma_1-\gamma_2)\left(\frac{\partial^4 v}{\partial x^4}\right)_{\gamma_1}\right] + (3-i)v_{21}\left[\frac{\gamma_2}{2(\gamma_2-\gamma_1)}\left(\frac{\partial v}{\partial x}\right)_{\gamma_1} + \right.$$
$$\left. + \frac{\gamma_2}{12}(\gamma_1-\gamma_2)\left(\frac{\partial^3 v}{\partial x^3}\right)_{\gamma_1}\right] - \left[\frac{\gamma_2}{2(\gamma_1-\gamma_2)}\left(\frac{\partial v}{\partial t}\right)_{\gamma_1} + \frac{\gamma_2}{12}(\gamma_1-\gamma_2)\left(\frac{\partial^3 v}{\partial x^2 \partial t}\right)_{\gamma_1}\right]. \tag{36}$$

Thus, integral from $\partial p_i(y)/\partial x$ by the width of the liquid layers has the form:

$$\frac{\partial}{\partial x}\int_{\gamma_1}^{\gamma_2} p_i dy = \frac{v_{i1}}{\mathrm{Re}}\left[\frac{\partial^3 A_i}{\partial x^3} + \frac{\gamma_1+2\gamma_2}{2\gamma_2^3}(\gamma_1^2-\gamma_2^2)\left(\frac{\partial c_1}{\partial x} + \gamma_2\frac{\partial c_2}{\partial x}\right)\right] + (3-i)v_{21}\frac{\partial^2 A_i}{\partial x^2} - \frac{\partial^2 A_i}{\partial x \partial t} + (\gamma_1-\gamma_2)\frac{\partial p_i}{\partial x}, \tag{37}$$

where $A_i=\frac{\gamma_2^2-\gamma_1^2}{2}\left\{\left[1+\frac{(\gamma_1+4\gamma_2)(\gamma_2^2-\gamma_1^2)}{20\gamma_2^3}\right]c_1+\left[\frac{\gamma_2-\gamma_1}{3}+\frac{(\gamma_1+4\gamma_2)(\gamma_2^2-\gamma_1^2)}{20\gamma_2^2}\right]c_2\right\}$, $\gamma_1$ and $\gamma_2$ depend on $i$=2,3, as $c_1$, $c_2$.

The expression (43) thus obtained is cumbersome, therefore a linear interpolation of a pressure may be used in liquid layers. Because $(v_2)_{h_2}=\partial q_2/\partial x$, $(v_3)_{h_1}=-\partial q_3/\partial x$, the momentum equations can be used by $y$ at the boundaries of the media:

$$\frac{\partial^2 q_i}{\partial t\partial x}+(3-i)v_{21}\frac{\partial^2 q_i}{\partial x^2}=-\frac{\partial p_i}{\partial y}+\frac{2v_{i1}}{\text{Re}}\frac{\partial^3 q_i}{\partial x^3},\quad \left(\frac{\partial p_i}{\partial y}\right)_{\gamma_i}=-\left[\frac{\partial^2 q_i}{\partial t\partial x}+v_{21}(3-i)\frac{\partial^2 q_i}{\partial x^2}-\frac{2v_{i1}}{\text{Re}}\frac{\partial^3 q_i}{\partial x^3}\right]. \tag{38}$$

Then expanding $p_i$ in a Taylor series by $y$ with accuracy to the linear terms: $p_i=(p_i)_{\gamma_1}+\left(\partial p_i/\partial y\right)_{\gamma_i}(y-\gamma_1)$, the searching integral can be computed in a form:

$$\frac{\partial}{\partial x}\int_{\gamma}^{\gamma_2} p_i dy=\frac{\partial}{\partial x}\left[-(p_i)_{\gamma_1}\gamma_1-\left(\frac{\partial p_i}{\partial y}\right)_{\gamma_i}\left(\frac{\gamma_1}{2}-\gamma_1^2\right)\right]=\frac{1}{2}\frac{\partial}{\partial x}\left(\frac{\partial p_i}{\partial y}\right)_{\gamma_i}-\left(\frac{\partial p_i}{\partial x}\right)_{\gamma_i}. \tag{39}$$

The terms of higher than the first order by perturbations were neglected in (39). Substituting (38) in (39):

$$\frac{\partial}{\partial x}\int_{\gamma_1}^{\gamma_2} p_i dy=\frac{1}{2}\left[\frac{2v_{i1}}{\text{Re}}\frac{\partial^4 q_1}{\partial x^4}+(i-3)v_{21}\frac{\partial^3 q_i}{\partial x^3}-\frac{\partial^3 q_i}{\partial t\partial x^2}\right]-\left(\frac{\partial p_i}{\partial x}\right)_{\gamma_i}, \tag{40}$$

where the pressure gradient by $x$ is taken from (13), (14). Condition (40) is simpler than (37) but it is not so precise. Integrating (10) for the vertical velocity component, with account of the profiles (26), yields the following pressure distribution in a liquid layers:

$$p_i=\frac{v_{i1}}{\text{Re}}\left[\frac{\partial^2 D}{\partial x^2}+d_2\left(\frac{1}{\gamma_2^2}-\frac{1}{y^2}\right)+2d_3\left(\frac{1}{\gamma_2^3}-\frac{1}{y^3}\right)+3d_4\left(\frac{1}{\gamma_2^4}-\frac{1}{y^4}\right)\right]+(3-i)v_{21}\frac{\partial D}{\partial x}+\frac{\partial D}{\partial t},\quad i=2,3, \tag{41}$$

$$D=d_1(y-\gamma_2)+d_2\ln\frac{y}{\gamma_2}+d_3\left(\frac{1}{\gamma_2}-\frac{1}{y}\right)+\frac{d_4}{2}\left(\frac{1}{\gamma_2^2}-\frac{1}{y^2}\right), \tag{42}$$

and then:

$$\frac{\partial}{\partial x}\int_{\gamma_1}^{\gamma_2} p_i dy=\frac{v_{i1}}{\text{Re}}\left[\frac{\partial^3 G}{\partial x^3}+\left(\frac{2}{\gamma_2}-\frac{\gamma_1}{\gamma_2^2}-\frac{1}{\gamma_1}\right)\frac{\partial d_2}{\partial x}+\left(\frac{3}{\gamma_2^2}-\frac{2\gamma_1}{\gamma_2^3}-\frac{1}{\gamma_1^2}\right)\frac{\partial d_3}{\partial x}+\right.$$
$$\left.+\left(\frac{4}{\gamma_2^3}-\frac{3\gamma_1}{\gamma_2^4}-\frac{1}{\gamma_1^2}\right)\frac{\partial d_4}{\partial x}\right]+(3-i)v_{21}\frac{\partial^2 G}{\partial x^2}-\frac{\partial^2 G}{\partial x\partial t}, \tag{43}$$

where $G=\gamma_1(\gamma_2-\gamma_1)d_1+\left(\gamma_1+\frac{\gamma_1^2-\gamma_2^2}{2\gamma_2}\right)d_2+\left(\frac{\gamma_2-\gamma_1}{\gamma_2}+\ln\frac{\gamma_1}{\gamma_2}\right)d_3+\left(1-\frac{\gamma_1^2+\gamma_2^2}{2\gamma_1\gamma_2}\right)\frac{d_4}{\gamma_2}$. Here the linear expansion into a Fourier series was performed: $\ln(y/\gamma_2)\approx(y-\gamma_2)/\gamma_2$ ($\int\ln x dx=x(\ln x-1)$).

Later on, the concrete expressions for each of liquids are analyzed. For example, from на основе (37) with account of (36), (38) in a linear approach:

$$\frac{\partial}{\partial x}\int_{1+\varsigma_2}^{\beta_2} p_2 dy = \frac{\nu_{21}}{\mathrm{Re}}\left[\frac{\partial^3 A_2}{\partial x^3} + \frac{1+2\beta_2}{2\beta_2^3}(1-\beta_2^2)\left(\frac{\partial c_{21}}{\partial x} + \beta_2\frac{\partial c_{22}}{\partial x}\right)\right] + \nu_{21}\frac{\partial^2 A_2}{\partial x^2} - \frac{\partial^2 A_2}{\partial x \partial t} + (\beta_2 - 1)\frac{\partial p_2}{\partial x}, \tag{44}$$

where are

$$A_2 = \frac{\beta_2^2 - 1}{2}\left\{\left[1 + \frac{(1+4\beta_2)(1-\beta_2^2)}{20\beta_2^3}\right]c_{21} + \left[\frac{\beta_2 - 1}{3} + \frac{(1+4\beta_2)(1-\beta_2^2)}{20\beta_2^2}\right]c_{22}\right\}, \tag{45}$$

$$p_2 = \frac{\beta_2}{2}\left\{\frac{\nu_{21}}{\mathrm{Re}}\left[\frac{1-\beta_2}{6}(v_2''')_{1+\varsigma_2} - (v_2'')_{1+\varsigma_2}\right] + \nu_{21}\left[\frac{(v_2')_{1+\varsigma_2}}{\beta_2 - 1} + \frac{1-\beta_2}{6}(v_2''')_{1+\varsigma_2}\right] + \frac{(\dot{v})_{1+\varsigma_2}}{1-\beta_2} + \frac{\beta_2 - 1}{6}(\dot{v}'')_{1+\varsigma_2}\right\},$$

where dash means differentiation by $x$, dot – derivative by $t$. For the third layer, respectively:

$$-\frac{\partial}{\partial x}\int_{-\beta_1}^{\varsigma_1} p_3 dy = \frac{\nu_{31}}{\mathrm{Re}}\left(\frac{\partial^3 A_3}{\partial x^3} + \frac{\partial c_{31}}{\partial x} - \beta_1\frac{\partial c_{32}}{\partial x}\right) - \frac{\partial^2 A_3}{\partial x \partial t} - \beta_1\frac{\partial p_3}{\partial x}, \tag{46}$$

$$p_3 = \frac{\beta_1 \nu_{31}}{2\,\mathrm{Re}}\left[(v_3'')_{\varsigma_1} - \frac{\beta_1}{6}(v_3''')_{\varsigma 1}\right] - \frac{1}{2}\left[(\dot{v}_3)_{\varsigma_1} - \frac{\beta_1^2}{6}(\dot{v}_3'')_{\varsigma_1}\right], \qquad A_3 = \frac{\beta_1^2}{5}\left(3c_{31} - \frac{\beta_1}{3}c_{32}\right). \tag{47}$$

If the liquid layers are thick enough ($\beta_i$ >> 1), then the following estimations may be got:

$$\frac{\partial}{\partial x}\int_{1+\varsigma_2}^{\beta_2} p_2 dy = \frac{\beta_2^2}{3}\left[\frac{\nu_{21}}{\mathrm{Re}_2}\left(\frac{v_2'''}{2} - \frac{\beta_2^2}{15}v_2^{V}\right) + \nu_{21}\left(v_2'' - \frac{\beta_2^2}{15}v_2^{IV}\right) + \frac{\beta_2^2}{15}\dot{v}_2''' - \dot{v}_2'\right], \tag{48}$$

$$-\frac{\partial}{\partial x}\int_{-\beta_1}^{\varsigma_1} p_3 dy = \frac{\beta_1^2}{15}\left(\frac{\beta_1^2}{3}\dot{v}_3''' - 8v_3'\right) - \frac{\beta_1^2 \nu_{31}}{5\,\mathrm{Re}}\left(\frac{3}{2}v_3''' + \frac{\beta_1^2}{9}v_3^{V}\right).$$

## Differential equations for the thick liquid layers

Substituting the approximate expressions (48) into (21) - (23), the following linear system of the differential equations is got for the perturbations (it is correct by $\beta_{2,3}$ >> 1, when liquid layers are substantially thick comparing to a gas layer):

$$\frac{\partial q_1}{\partial x} = \frac{\partial \varsigma_1}{\partial t} + \frac{\partial \varsigma_1}{\partial x} - \frac{\partial \varsigma_2}{\partial t} + (\nu_{21} - 1)\frac{\partial \varsigma_2}{\partial x}, \quad \frac{\partial q_1}{\partial t} + \frac{\partial q_1}{\partial x} = -\frac{\partial p_1}{\partial x} + \frac{2}{\mathrm{Re}}\left[\frac{\partial^2 \varsigma_1}{\partial t \partial x} + \frac{\partial^2 \varsigma_1}{\partial x^2} - \frac{\partial^2 \varsigma_2}{\partial t \partial x} + (\nu_{21} - 1)\frac{\partial^2 \varsigma_2}{\partial x^2}\right]; \tag{49}$$

$$\frac{\partial q_2}{\partial x} = \frac{\partial \varsigma_2}{\partial t} + (1 - \nu_{21})\frac{\partial \varsigma_2}{\partial x}, \tag{50}$$

$$\frac{\partial q_2}{\partial t} = \nu_{21}\frac{\partial q_2}{\partial x} + \frac{\beta_2^2}{3}\left[\frac{\nu_{21}}{\mathrm{Re}}\left(\frac{\beta_2^2}{15}\frac{\partial^6 q_2}{\partial x^6} - \frac{1}{2}\frac{\partial^4 q_2}{\partial x^4}\right) + \nu_{21}\left(\frac{\beta_2^2}{15}\frac{\partial^5 q_2}{\partial x^5} - \frac{\partial^3 q_2}{\partial x^3}\right) + \frac{\partial^5 q_2}{\partial t \partial x} - \frac{\beta_2^2}{15}\frac{\partial^5 q_2}{\partial t \partial x^4}\right] + \frac{2\nu_{21}}{\mathrm{Re}}\frac{\partial^2 q_2}{\partial x^2};$$

$$\frac{\partial q_3}{\partial x} = -\frac{\partial \varsigma_1}{\partial t} - \frac{\partial \varsigma_1}{\partial x}, \qquad \frac{\partial q_3}{\partial t} = \frac{\beta_1^2}{15}\left(\frac{\beta_1^2}{3}\frac{\partial^5 q_3}{\partial t \partial x^4} - 8\frac{\partial^3 q_3}{\partial t \partial x^2}\right) - \frac{\beta_1^2 \nu_{31}}{5\,\mathrm{Re}}\left(\frac{3}{2}\frac{\partial^4 q_3}{\partial x^4} + \frac{\beta_1^2}{9}\frac{\partial^6 q_3}{\partial x^6}\right) + \frac{2\nu_{31}}{\mathrm{Re}}\frac{\partial^2 q_3}{\partial x^2}. \tag{51}$$

The kinematic boundary conditions (13), (14) were used.

Continuing with application of the polynomial approximations (26), the gradients of the transversal velocities on the boundaries are computed and substituted into the dynamical boundary conditions (13), (14), which result in $p_1$ for further use in (49). It gives additional condition for the perturbations. First the gradients are computed with (26) and (30)-(32) for the case $\beta_{2,3} >> 1$:

$$\left(\frac{\partial v_1}{\partial y}\right)_{\varsigma_1} = \frac{\partial \varsigma_2}{\partial t} + (1-v_{21})\frac{\partial \varsigma_2}{\partial x} - \frac{\partial \varsigma_1}{\partial t} - \frac{\partial \varsigma_1}{\partial x} + \frac{1}{6}\left[\frac{\partial^3 \varsigma_2}{\partial t \partial x^2} + (1-v_{21})\frac{\partial^3 \varsigma_2}{\partial x^3}\right] - \frac{2}{3}\left(\frac{\partial^3 \varsigma_1}{\partial t \partial x^2} + \frac{\partial^3 \varsigma_1}{\partial x^3}\right),$$

$$\left(\frac{\partial v_1}{\partial y}\right)_{1+\varsigma_2} = \frac{\partial \varsigma_2}{\partial t} + (1-v_{21})\frac{\partial \varsigma_2}{\partial x} + \frac{1}{2}\left(\frac{\partial^3 \varsigma_1}{\partial t \partial x^2} + \frac{\partial^3 \varsigma_1}{\partial x^3}\right) - \frac{\partial \varsigma_1}{\partial t} - \frac{\partial \varsigma_1}{\partial x}, \quad \left(\frac{\partial v_3}{\partial y}\right)_{\varsigma_1} = \frac{1}{\beta_1}\left(\frac{\partial \varsigma_1}{\partial t} + \frac{\partial \varsigma_1}{\partial x}\right) + \frac{\beta_1}{3}\left(\frac{\partial^3 \varsigma_1}{\partial t \partial x^2} + \frac{\partial^3 \varsigma_1}{\partial x^3}\right),$$

$$\left(\frac{\partial v_2}{\partial y}\right)_{1+\varsigma_2} = -\frac{1}{\beta_2}\left[\frac{\partial \varsigma_2}{\partial t} + (1-v_{21})\frac{\partial \varsigma_2}{\partial x}\right] - \frac{\beta_2}{3}\left[\frac{\partial^3 \varsigma_2}{\partial t \partial x^2} + (1-v_{21})\frac{\partial^3 \varsigma_2}{\partial x^3}\right]. \qquad (52)$$

### Determination of the pressure at the boundaries of liquid layers

The correlations (34), (36) and (25) are used in a linear approach with estimation $\beta_{2,3} >> 1$ (all terms of order 1 are small comparing to $\beta_{2,3}$, etc.):

$$(p_2)_{1+\varsigma_2} = \frac{\beta_2^2}{12}\left[\frac{\partial^4 \varsigma_2}{\partial t^2 \partial x^2} + (1-2v_{21})\frac{\partial^4 \varsigma_2}{\partial t \partial x^3} + v_{21}(v_{21}-1)\frac{\partial^4 \varsigma_2}{\partial x^4}\right] + \qquad (53)$$

$$+\frac{3}{2}\left[v_{21}(1-v_{21})\frac{\partial^2 \varsigma_2}{\partial x^2} + (2v_{21}-1)\frac{\partial^2 \varsigma_2}{\partial t \partial x} - \frac{\partial^2 \varsigma_2}{\partial t^2}\right] - \frac{\beta_2 v_{21}}{2\,\mathrm{Re}}\left[\frac{\partial^3 \varsigma_2}{\partial t \partial x^2} + (1-v_{21})\frac{\partial^3 \varsigma_2}{\partial x^3} + \frac{\beta^2}{6}\left(\frac{\partial^5 \varsigma_2}{\partial t \partial x^4} + (1-v_{21})\frac{\partial^5 \varsigma_2}{\partial x^5}\right)\right],$$

$$(p_3)_{\varsigma_1} = \frac{\beta_1 v_{31}}{2\,\mathrm{Re}}\left[\frac{\partial^3 \varsigma_1}{\partial t \partial x^2} + \frac{\partial^3 \varsigma_1}{\partial x^3} - \frac{\beta_1}{6}\left(\frac{\partial^5 \varsigma_1}{\partial t \partial x^4} + \frac{\partial^5 \varsigma_1}{\partial x^5}\right)\right] + \frac{1}{2}\left[\frac{\beta_1^2}{6}\left(\frac{\partial^4 \varsigma_1}{\partial t^2 \partial x^2} + \frac{\partial^4 \varsigma_1}{\partial t \partial x^3}\right) - \left(\frac{\partial^2 \varsigma_1}{\partial t^2} + \frac{\partial^2 \varsigma_1}{\partial t \partial x}\right)\right].$$

All mixed derivatives by $x$ and $t$ are adopted here. And then substituting the obtained correlations (52), (53) into the boundary conditions (13), (14) for the pressure results

$$p_1 = \frac{1}{We_3}\frac{\partial \varsigma_1}{\partial x^2} + \frac{2}{\mathrm{Re}}\left\{\frac{\partial \varsigma_2}{\partial t} + (1-v_{21})\frac{\partial \varsigma_2}{\partial x} + \frac{1}{6}\left[\frac{\partial^3 \varsigma_2}{\partial t \partial x^2} + (1-v_{21})\frac{\partial^3 \varsigma_2}{\partial x^3}\right] - \right.$$

$$\left. -\left(1+\frac{\mu_{31}}{\beta_1}\right)\left(\frac{\partial \varsigma_1}{\partial t} + \frac{\partial \varsigma_1}{\partial x}\right)\right\} + \frac{\rho_{31}}{2}\left[\frac{\beta_1^2}{6}\left(\frac{\partial^4 \varsigma_1}{\partial t^2 \partial x^2} + \frac{\partial^4 \varsigma_1}{\partial t \partial x^3}\right) - \left(\frac{\partial^2 \varsigma_1}{\partial t^2} + \frac{\partial^2 \varsigma_1}{\partial x \partial t}\right)\right] -$$

$$-\frac{\mu_{31}\beta_1 + 8}{6\,\mathrm{Re}}\left(\frac{\partial^3 \varsigma_1}{\partial t \partial x^2} + \frac{\partial^3 \varsigma_1}{\partial x^3}\right) - \frac{\rho_{31}\beta_1^2}{12\,\mathrm{Re}}v_{31}\left(\frac{\partial^5 \varsigma_1}{\partial t \partial x^4} + \frac{\partial^5 \varsigma_1}{\partial x^5}\right), \qquad (54)$$

$$p_1 = -\frac{1}{We_2}\frac{\partial^2 \varsigma_2}{\partial x^2} + \frac{2}{\mathrm{Re}}\left\{\left(1+\frac{\mu_{21}}{\beta_2}\right)\left[\frac{\partial \varsigma_2}{\partial t} + (1-v_{21})\frac{\partial \varsigma_2}{\partial x}\right] + \frac{1}{2}\left(\frac{\partial^3 \varsigma_1}{\partial t \partial x^2} + \frac{\partial^3 \varsigma_1}{\partial x^3}\right) - \frac{\partial \varsigma_1}{\partial t} - \frac{\partial \varsigma_1}{\partial x}\right\} +$$

$$+\frac{\mu_{21}\beta_2}{6\,\mathrm{Re}}\left[\frac{\partial^3 \varsigma_2}{\partial t \partial x^2} + (1-v_{21})\frac{\partial^3 \varsigma_2}{\partial t \partial x^3}\right] + \rho_{21}\left\{\frac{\beta_2^2}{12}\left[\frac{\partial^4 \varsigma_2}{\partial t^2 \partial x^2} + (1-2v_{21})\frac{\partial^4 \varsigma_2}{\partial t \partial x^3} + v_{21}(v_{21}-1)\frac{\partial^4 \varsigma_2}{\partial x^4}\right] + \right.$$

$$+\frac{3}{2}\left[\nu_{21}(1-\nu_{21})\frac{\partial^2\varsigma_2}{\partial x^2}+(2\nu_{21}-1)\frac{\partial^2\varsigma_2}{\partial t\partial x}-\frac{\partial^2\varsigma_2}{\partial t^2}\right]-\frac{\beta_2^2\nu_{21}}{12\,\mathrm{Re}}\left[\frac{\partial^5\varsigma_2}{\partial t\partial x^4}+(1-\nu_{21})\frac{\partial^5\varsigma_2}{\partial x^5}\right]\Bigg]\Bigg\}.$$

## 4. Differential equations for perturbations of the gas layer boundaries

The differential equation for the perturbations of the boundaries of layers is derived using (54). Based on the (54), from (49) - (51), the following three equations for the boundaries' perturbations are obtained:

$$\begin{aligned}&\frac{\partial^2\varsigma_1}{\partial t^2}+2\frac{\partial^2\varsigma_1}{\partial t\partial x}+\frac{\partial^2\varsigma_1}{\partial x^2}-\frac{\partial^2\varsigma_2}{\partial t^2}+(\nu_{21}-2)\frac{\partial^2\varsigma_2}{\partial t\partial x}+(\nu_{21}-1)\frac{\partial^2\varsigma_2}{\partial x^2}-\frac{1}{We_2}\frac{\partial^4\varsigma_2}{\partial x^4}+\\&+\frac{2}{\mathrm{Re}}\left[\frac{1}{2}\left(\frac{\partial^5\varsigma_1}{\partial t\partial x^4}+\frac{\partial^5\varsigma_1}{\partial x^5}\right)+\left(1+\frac{\mu_{21}}{\beta_2}\right)\left(\frac{\partial^3\varsigma_2}{\partial t\partial x^2}+(1-\nu_{21})\frac{\partial^3\varsigma_2}{\partial x^3}\right)-\frac{\partial^3\varsigma_1}{\partial x^2 t}-\frac{\partial^3\varsigma_1}{\partial x^3}\right]+\\&+\frac{\mu_{21}\beta_2}{6\,\mathrm{Re}}\left[\frac{\partial^5\varsigma_2}{\partial t\partial x^4}+(1-\nu_{21})\frac{\partial^5\varsigma_2}{\partial x^5}\right]+\beta_{21}\left\{\frac{\beta_2^2}{12}\left[\nu_{21}(1-\nu_{21})\frac{\partial^6\varsigma_2}{\partial x^6}+(2\nu_{21}-1)\frac{\partial^6\varsigma_2}{\partial t\partial x^5}+\frac{\partial^6\varsigma_2}{\partial t^2\partial x^4}\right]+\right.\\&\left.+\frac{3}{2}\left[\frac{\partial^4\varsigma_2}{\partial t^2\partial x^2}+(1-2\nu_{21})\frac{\partial^4\varsigma_2}{\partial t\partial x^3}+\nu_{21}(\nu_{21}-1)\frac{\partial^4\varsigma_2}{\partial x^4}\right]+\frac{\beta_2^2\nu_{21}}{12\,\mathrm{Re}}\left[\frac{\partial^7\varsigma_2}{\partial t\partial x^6}+(1-\nu_{21})\frac{\partial^7\varsigma_2}{\partial x^7}\right]\right\}=0,\end{aligned}\tag{55}$$

$$\begin{aligned}&\frac{1}{We_3}\frac{\partial^2\varsigma_1}{\partial x^2}+\frac{1}{We_2}\frac{\partial^2\varsigma_2}{\partial x^2}+\frac{\rho_{31}}{2}\left[\frac{\beta_1^2}{6}\left(\frac{\partial^4\varsigma_1}{\partial t^2\partial x^2}+\frac{\partial^4}{\partial t\partial x^3}\right)-\frac{\partial^2\varsigma_1}{\partial t^2}-\frac{\partial^2\varsigma_1}{\partial t\partial x}\right]-\\&-\frac{1}{12\,\mathrm{Re}}\left[2(\mu_{31}\beta_1+14)\left(\frac{\partial^3\varsigma_1}{\partial t\partial x^2}+\frac{\partial^3\varsigma_1}{\partial x^3}\right)+\mu_{31}\beta_1^2\left(\frac{\partial^5\varsigma_1}{\partial t\partial x^4}+\frac{\partial^5\varsigma_1}{\partial x^5}\right)\right]+\\&+\frac{2}{\mathrm{Re}}\left\{\frac{\mu_{21}}{\beta_2}\left[(\nu_{21}-1)\frac{\partial\varsigma_2}{\partial x}-\frac{\partial\varsigma_2}{\partial t}\right]-\frac{\mu_{21}}{\beta_1}\left(\frac{\partial\varsigma_1}{\partial t}+\frac{\partial\varsigma_1}{\partial x}\right)+\frac{1}{6}\left[\frac{\partial^3\varsigma_2}{\partial t\partial x^2}+(1-\nu_{21})\frac{\partial^3\varsigma_2}{\partial x^3}\right]\right\}+\\&+\frac{\mu_{21}\beta_2}{6\,\mathrm{Re}}\left[(\nu_{21}-1)\frac{\partial^3\varsigma_2}{\partial x^3}-\frac{\partial^3\varsigma_2}{\partial t\partial x^2}\right]+\rho_{21}\left\{\frac{\beta_2^2}{12}\left[\nu_{21}(1-\nu_{21})\frac{\partial^4\varsigma_2}{\partial x^4}+2(\nu_{21}-1)\frac{\partial^4\varsigma_2}{\partial t\partial x^3}-\frac{\partial^4\varsigma_2}{\partial t^2\partial x^2}\right]+\right.\\&\left.+\frac{3}{2}\left[\frac{\partial^2\varsigma_2}{\partial t^2}+(1-2\nu_{21})\frac{\partial^2\varsigma_2}{\partial t\partial x}+\nu_{21}(\nu_{21}-1)\frac{\partial^2\varsigma_2}{\partial x^2}\right]+\frac{\beta_2^2\nu_{21}}{12\,\mathrm{Re}}\left[\frac{\partial^5\varsigma_2}{\partial t\partial x^4}+(1-\nu_{21})\frac{\partial^5\varsigma_2}{\partial x^5}\right]\right\}=0,\end{aligned}\tag{56}$$

$$\begin{aligned}&\frac{\partial^2\varsigma_2}{\partial t^2}+(1-2\nu_{21})\frac{\partial^2\varsigma_2}{\partial t\partial x}+\nu_{21}(\nu_{21}-1)\frac{\partial^2\varsigma_2}{\partial x^2}+\frac{\beta_2^2}{3}\left\{\frac{\nu_{21}}{\mathrm{Re}}\left[\frac{\beta_2^2}{15}\left((\nu_{21}-1)\frac{\partial^8\varsigma_2}{\partial x^8}-\frac{\partial^8\varsigma_2}{\partial t\partial x^7}\right)+\right.\right.\\&\left.+\frac{1}{2}\left(\frac{\partial^6\varsigma_2}{\partial t\partial x^5}+(1-\nu_{21})\frac{\partial^6\varsigma_2}{\partial x^6}\right)\right]+\nu_{21}\left[\frac{\beta_2^2}{15}(\nu_{21}-1)\frac{\partial^7\varsigma_2}{\partial x^7}+\frac{\partial^7\varsigma_2}{\partial t\partial x^6}+(1-\nu_{21})\frac{\partial^5\varsigma_2}{\partial x^5}\right]-\\&\left.-\frac{\partial^4\varsigma_2}{\partial t^2\partial x^2}+\frac{\beta_2^2}{15}(1-2\nu_{21})\frac{\partial^7\varsigma_2}{\partial t\partial x^6}+\frac{\beta_2^2}{15}\frac{\partial^7\varsigma_2}{\partial t^2\partial x^5}\right\}+\left[\frac{\beta_2^2}{3}(\nu_{21}-1)-\frac{2\nu_{21}}{\mathrm{Re}}\right]\frac{\partial^4\varsigma_2}{\partial t\partial x^3}+\frac{2\nu_{21}}{\mathrm{Re}}(\nu_{21}-1)\frac{\partial^4\varsigma_2}{\partial x^4}=0,\end{aligned}\tag{57}$$

$$\begin{aligned}&\frac{\partial^2\varsigma_1}{\partial t^2}+\frac{\partial^2\varsigma_1}{\partial t\partial x}+\frac{\beta_1^2}{15}\left[8\left(\frac{\partial^5\varsigma_1}{\partial t^2\partial x^3}+\frac{\partial^5\varsigma_1}{\partial t\partial x^4}\right)-\frac{\beta_1^2}{3}\left(\frac{\partial^7\varsigma_1}{\partial t^2\partial x^5}+\frac{\partial^7\varsigma_1}{\partial t\partial x^6}\right)\right]+\\&+\frac{\beta_1^2}{5\,\mathrm{Re}_3}\left[\frac{3}{2}\left(\frac{\partial^6\varsigma_1}{\partial t\partial x^5}+\frac{\partial^6\varsigma_1}{\partial x^6}\right)+\frac{\beta_1^2}{9}\left(\frac{\partial^8\varsigma_1}{\partial t\partial x^7}+\frac{\partial^8\varsigma_1}{\partial x^8}\right)\right]-\frac{2}{\mathrm{Re}_3}\left(\frac{\partial^4\varsigma_1}{\partial t\partial x^3}+\frac{\partial^4\varsigma_1}{\partial x^4}\right)=0.\end{aligned}\tag{58}$$

Analysis of the system (55) - (58) allows studying the stability of the boundaries of the gas flow. The partial differential equation array includes the derivatives of a higher order up to the eighth order, therefore it difficult for solving. Four equations totally, two functions sought, which is a consequence of the approximations applied for the profiles of transversal velocities and pressures of the liquid layers. As far as two last equations are autonomous, they can be solved independently, considering for example the simple harmonic waves in the form $\varsigma_i = x_j e^{i(k_j x - \omega_j t + \phi_j)}$, where $j$=1,2, $i = \sqrt{-1}$, $x_j$ – constants, the initial amplitudes of the perturbations, $k_j$, $\omega_j$- the wave numbers and frequencies of the oscillations, $\varphi_j$ – the initial phases of perturbations. Afterward, substituting the obtained solution into the other two equations of the system (58), we can get dispersion equations for computing the frequencies of perturbations $\omega_j = \omega_j\,(k_j)$ depending on the wave numbers.

The perturbations of the top and lower boundaries are interconnected and can differ only by the initial phases $\varphi_j$, $k_1 = k_2 = k$, $\omega_1 = \omega_2 = \omega$:

$$e^{i(k_1 x - \omega_1 t + \varphi_1)} = const\; e^{i(k_2 x - \omega_2 t + \varphi_2)}, \tag{59}$$

With account of the above, substituting (59) into (57), (58) after the contraction of the exponent and the amplitude $x_j$ (the equations are linear homogeneous in terms of the perturbations), we obtain:

$$\left[\frac{\beta_1^2}{15}k^3 i\left(\frac{\beta_1^2 k^2}{3}+8\right)-1\right]\omega^2+\left[1-\frac{k^3}{15}\beta_1^2 i\left(\frac{\beta_1^2 k^2}{3}+8\right)+\frac{k^2\nu_{31}}{\mathrm{Re}}\left(\frac{3}{10}\beta_1^2 k^2-\frac{k^4}{45}\beta_1^4+2\right)\right]k\omega+ \\ +\frac{k^4\nu_{31}}{\mathrm{Re}}\left(\frac{k^4}{45}\beta_1^4-\frac{3}{10}\beta_1^2 k^2-2\right)=0, \tag{60}$$

$$\left(1+\frac{\beta_2^2}{3}k^2+\frac{\beta_2^4}{45}k^5 i\right)\omega^2+\left\{\frac{\beta_2^2\nu_{31}}{3\mathrm{Re}}\left(\frac{\beta_2^2}{15}k^2-\frac{1}{2}\right)k^4+\frac{\beta_2^2}{3}\left[\frac{\beta_2^2}{15}(2\nu_{21}-1)-\nu_{21}\right]ik^5+ \right. \\ \left. +\left[\frac{\beta_2^2}{3}(\nu_{21}-1)-\frac{2\nu_{21}}{\mathrm{Re}}\right]k^3+2\nu_{21}-1\right\}k\omega+(\nu_{21}-1)k^2\left\{\nu_{21}-\frac{2k^2}{\mathrm{Re}_2}+ \right. \\ \left. +\frac{\beta_2^2}{3}\left[\nu_{21}\left(\frac{\beta_2^2}{15}k^2+1\right)\right]k^3 i-\frac{k^4\nu_{21}}{\mathrm{Re}}\left(\frac{\beta_2^2}{15}k^2+\frac{1}{2}\right)\right]\right\}=0. \tag{61}$$

The reasonable way for further solution of the problem seems to be as follows. Starting with the velocity approximation (26), after integration of (10) we can get the pressure distribution (41). Then integrating (41) with account linear approach $\frac{\partial}{\partial x}\int_{\gamma_1}^{\gamma_2} p_i dy = \int_{\gamma_1}^{\gamma_2}\frac{\partial p_i}{\partial x}dy$, as far as product $\frac{\partial \gamma_i}{\partial x}$ and $p_i$ are of the second order, yields:

$$\int_{\gamma_1}^{\gamma_2}\frac{\partial p_i}{\partial x}dy,\quad \frac{\partial}{\partial x}\int_{\gamma_1}^{\gamma_2} p_i dy = \frac{\nu_{i1}}{\mathrm{Re}}\left(\frac{\partial^3 G}{\partial x^3}+\frac{\partial F}{\partial x}\right)+(3-i)\nu_{21}\frac{\partial^2 G}{\partial x^2}+\frac{\partial^2 G}{\partial t\partial x}+(\gamma_2-\gamma_1)\frac{\partial (p_i)_{\gamma_1}}{\partial x} \tag{62}$$

where a linearization must be done too. Here

$$G = -\frac{d_1}{2}(\gamma_1-\gamma_2)^2 + d_2\left(\gamma_1-\gamma_2+\gamma_1\ln\frac{\gamma_2}{\gamma_1}\right)+d_3\left(\ln\frac{\gamma_1}{\gamma_2}-\frac{\gamma_1}{\gamma_2}+1\right)-d_4\frac{(\gamma_1-\gamma_2)^2}{2\gamma_1^2\gamma_2},$$

$$F = d_2\left(-\frac{\gamma_1}{\gamma_2^2}+\frac{2}{\gamma_2}-\frac{1}{\gamma_1}\right)+2d_3\left(-\frac{\gamma_1}{\gamma_2^3}-\frac{1}{2\gamma_1^2}+\frac{3}{2\gamma_2^2}\right)+3d_4\left(-\frac{\gamma_1}{\gamma_2^4}-\frac{1}{3\gamma_1^3}+\frac{4}{3\gamma_2^3}\right), \qquad v_i = d_1+\frac{d_2}{y}+\frac{d_3}{y^2}+\frac{d_4}{y^3},$$

$$\frac{\partial v_i}{\partial y} = -\frac{1}{y^2}\left(d_2+2\frac{d_3}{y}+3\frac{d_4}{y^2}\right), \ (p_i)_{\gamma_1} = \rho_{1i}\left(p_1-\frac{1}{We_i}\frac{\partial^2\varsigma_i}{\partial x^2}+2\frac{\mu_{i1}}{\mathrm{Re}}\frac{\partial v_i}{\partial y}\right)_{\gamma_1} \text{, from (13), (14).} \qquad (63)$$

Substituting (62) in (21)-(23) results a system for $\zeta_i$.

Pressure is obtained by integrating (10), accounting that by taking differential out of integral all terms with differentials from the limit of integration are as follows (second-order):

$$\frac{\partial}{\partial x}\int_{\gamma_1}^{y} f(y)dy = \int_{\gamma_1}^{y}\frac{df}{dx}dy+(f)_y\frac{\partial y}{\partial x}-(f)_{\gamma_1}\frac{\partial \gamma_1}{\partial x}$$

and thus $\frac{\partial}{\partial x}\int_{\gamma_1}^{y} f(y)dy = \int_{\gamma_1}^{y}\frac{df}{dx}dy$, $\frac{\partial^2}{\partial x^2}\int_{\gamma_1}^{y} f(y)dy = \int_{\gamma_1}^{y}\frac{d^2 f}{dx^2}dy$ and so on. Therefore

$$p_i = \frac{\nu_{i1}}{\mathrm{Re}}\left[\frac{\partial^2 D_i}{\partial x^2}+\left(\frac{\partial v_i}{\partial y}\right)_{\gamma_1}^{y}\right]+\nu_{21}(3-i)\frac{\partial D_i}{\partial x}-\frac{\partial D_i}{\partial t}+(p_i)_{\gamma_1}, \quad v_i = d_1+\frac{d_2}{y}+\frac{d_3}{y^2}+\frac{d_4}{y^3},$$

$$D_i = \int_{\gamma_1}^{y} v_i dy = d_1(y-\gamma_1)+d_2\ln\frac{y}{\gamma_1}+d_3\left(\frac{1}{\gamma_1}-\frac{1}{y}\right)+\frac{d_4}{2}\left(\frac{1}{\gamma_1^2}-\frac{1}{y^2}\right), \ \frac{\partial v_i}{\partial y} = -\frac{1}{y^2}\left(d_2+2\frac{d_3}{y}+3\frac{d_4}{y^2}\right),$$

$(p_i)_{\gamma 1}$ are from (44), (45). For (21)-(23) $\int_{\gamma_1}^{\gamma_2}\frac{\partial p_i}{\partial x}dy$ and 2 equations for $\zeta_i$.

$$\int_{\gamma_1}^{\gamma_2} p_i dy = \frac{\nu_{i1}}{\mathrm{Re}}\left[\frac{\partial^2 G_i}{\partial x^2}-(v_i)_{\gamma_i}-\left(\frac{\partial v_i}{\partial y}\right)_{\gamma_i}(\gamma_2-\gamma_1)\right]+\nu_{21}(3-i)\frac{\partial G_i}{\partial x}-\frac{\partial G_i}{\partial t}+(p_i)_{\gamma_1}(\gamma_2-\gamma_1),$$

$$G_i = \int_{\gamma_1}^{\gamma_2} D_i dy, \ G_i = \int_{\gamma_1}^{\gamma_2} D_i dy = \frac{d_1}{2}(\gamma_2-\gamma_1)^2+d_2\left(\gamma_2\ln\frac{\gamma_2}{\gamma_1}+\gamma_1-\gamma_2\right)+d_3\left(\frac{\gamma_2}{\gamma_1}+\ln\frac{\gamma_1}{\gamma_2}-1\right)+d_4\frac{(\gamma_2-\gamma_1)^2}{2\gamma_1^2\gamma_2}.$$

For upper boundary (44) results with account $\beta_2 >> 1$:

$$G_2 = 3\left(\frac{\beta_2}{2}-2\ln\beta_2\right)(v_2)_{1+\varsigma_2}+\frac{1}{2}\left(\ln\beta_2-\frac{\beta_2}{6}\right)\left(\frac{\partial^2 v_2}{\partial x^2}\right)_{1+\varsigma_2},$$

$$\frac{\partial}{\partial x}\int_{1+\varsigma_2}^{\beta_2} p_2 dy = \frac{\nu_{21}}{\mathrm{Re}}\left[\frac{\partial^3 G_2}{\partial x^3}-\left(\frac{\partial v_2}{\partial x}\right)_{1+\varsigma_2}-\beta_2\left(\frac{\partial v_2}{\partial x}+\frac{1}{6}\frac{\partial^3 v_2}{\partial x^3}\right)_{1+\varsigma_2}\right]+\nu_{21}\frac{\partial^2 G_2}{\partial x^2}-\frac{\partial^2 G_2}{\partial t\partial x}+\left(\frac{\partial p_2}{\partial x}\right)_{1+\varsigma_2}\beta_2,$$

$G_3$ has terms of the $2^{nd}$ order and higher $\Rightarrow$ $G_3 = 0$,

$$\frac{\partial}{\partial x}\int_{\varsigma_1}^{-\beta_1} p_3 dy = \frac{\nu_{31}}{\mathrm{Re}}\left[-(v_3')_{\varsigma_1}+\left(\frac{2}{\beta_1}-\frac{1}{\varsigma_1}\right)(v_3')_{\varsigma_1}(\beta_1+\varsigma_1)\right]+(p_3')_{\varsigma_1}(-\beta_1-\varsigma_1)\approx$$

$$\approx\beta_1\left[\frac{\nu_{31}}{\mathrm{Re}}\frac{\partial}{\partial x}\left(\frac{v_3}{\varsigma_1}\right)+\frac{\rho_{13}}{We_3}\frac{\partial^3\varsigma_1}{\partial x^3}-4\frac{\nu_{31}}{\mathrm{Re}\,\beta_1}\frac{\partial v_3}{\partial x}-\rho_{13}\frac{\partial p_1}{\partial x}\right], \qquad \left(\frac{\partial v_3}{\partial y}\right)_{\varsigma_1} = \left(\frac{2}{\beta_1}-\frac{1}{\varsigma_1}\right)(v_3)_{\varsigma_1} \text{ - zero order.}$$

And ($\beta_2 >> 1$):

$$\frac{\partial}{\partial x}\int_{1+\varsigma_2}^{\beta_2} p_2 dy = \frac{\nu_{21}}{\mathrm{Re}}\left[\frac{1}{2}\left(\ln\beta_2-\frac{\beta_2}{6}\right)v_2^{V}+\left(\beta_2-6\ln\beta_2\right)v_3'''-3\beta_2 v_2'\right]+$$

$$+\nu_{21}\left[3\left(\frac{\beta_2}{2}-2\ln\beta_2\right)v_2''+2\left(\ln\beta_2-\frac{\beta_2}{6}\right)v_v^{IV}\right]+3\left(2\ln\beta_2-\frac{\beta_2}{2}\right)\dot{v}_2'+$$

$$+\frac{1}{2}\left(\frac{\beta_2}{6}-\ln\beta_2\right)\dot{v}_2''''+\beta_2\rho_{12}\left(\frac{\partial p_1}{\partial x}-\frac{1}{We_2}\frac{\partial^3\varsigma_2}{\partial x^3}\right).$$

Substitution into (21)-(23) yields:

$$\frac{\partial q_1}{\partial x}=\frac{\partial\varsigma_1}{\partial t}+\frac{\partial\varsigma_1}{\partial x}-\frac{\partial\varsigma_2}{\partial t}+(\nu_{21}-1)\frac{\partial\varsigma_2}{\partial x},\quad \frac{\partial q_1}{\partial t}+\frac{\partial q_1}{\partial x}=-\frac{\partial p_1}{\partial x}-\frac{2}{\mathrm{Re}}\frac{\partial^2 q_1}{\partial x^2},\quad \frac{\partial q_2}{\partial x}=\frac{\partial\varsigma_2}{\partial t}+(1-\nu_{21})\frac{\partial\varsigma_2}{\partial x},$$

$$\frac{\partial q_2}{\partial t}-\nu_{21}\frac{\partial q_2}{\partial x}=\frac{\nu_{21}}{\mathrm{Re}}\left[3\beta_2 v_2'+\left(6\ln\beta_2-\beta_2\right)v_2''''+\frac{1}{2}\left(\frac{\beta_2}{6}-\ln\beta_2\right)v_2^{V}\right]+$$

$$+\nu_{21}\left[3\left(2\ln\beta_2-\frac{\beta_2}{2}\right)v_2''+2\left(\frac{\beta_2}{6}-\ln\beta_2\right)v_2^{IV}\right]+3\left(\frac{\beta_2}{2}-2\ln\beta_2\right)\dot{v}_2'+$$

$$+\frac{1}{2}\left(\ln\beta_2-\frac{\beta_2}{6}\right)\dot{v}_2''''+\beta_2\rho_{12}\left(\frac{1}{We_2}\frac{\partial^3\varsigma_2}{\partial x^3}-\frac{\partial p_1}{\partial x}\right),$$

$$\frac{\partial q_3}{\partial x}=-\left(\frac{\partial\varsigma_1}{\partial t}+\frac{\partial\varsigma_1}{\partial x}\right),\quad \frac{\partial q_3}{\partial x}=\beta_1\left[\frac{\nu_{31}}{\mathrm{Re}}\frac{\partial}{\partial x}\left(\frac{v_3}{\varsigma_1}\right)+\frac{\rho_{13}}{We_3}\frac{\partial^3\varsigma_1}{\partial x^3}-\rho_{13}\frac{\partial p_1}{\partial x}\right]-\frac{2\nu_{31}}{\mathrm{Re}}v_3'.$$

In the momentum equation for the second phase (upper liquid layer) the terms $\ln\beta_2$ are kept, because $\beta_2 >> 1$ can be by: $\ln\beta_2\sim\beta_2$ (e.g., $\beta_2=100$, $\ln\beta_2\approx 4{,}6$; $\beta_2=10$, $\ln\beta_2\approx 2{,}3$; $\beta_2=1000$, $\ln\beta_2\approx 6{,}9$; $\beta_2=10^4$, $\ln\beta_2=9{,}2$ obviously the terms with $\ln\beta_2\sim 1$, when $\beta_2\sim 10$, but $\ln\beta_2\sim 10$ for $\beta_2$ in a substantially wide range of $\beta_2$. Thus, the terms with $\ln\beta_2$ can be omitted when $\beta_2\sim 10$ and $\beta_2\sim 1000$ and higher, and around $\beta_2\sim 100$ they can be substantial and depending of specific values because they are multiplayers with bigger ones than $\beta_2$).

Further work must be done with computational experiment and analysis of the results obtained. The model thus derived may be useful in the investigations of some physical problems including stability of the vapor layer around the hot particle during its cooling in a volatile liquid, for revealing the peculiarities of the heat transfer critical heat flux.

## References

1. H.S. Park, I.V. Kazachkov, B.R. Sehgal, Y. Maruyama and S. Fujui, "Analysis of plunging jet penetration into liquid pool with various densities", Fourth International Conference on Multiphase Flow, New Orleans, Louisiana, USA, May 27-th June 1 (2001).
2. K.A. Bin, "Gas entrainment by plunging liguid jets", Chem. Eng. Sci. 48, 3585 (1993).
3. H. Chanson, *Air Bubble Entrainment in Free-Surface Turbulent Shear Flows.* (Academic Press, San Diego, CA, 1996).
4. H.S. Park, N. Yamano, K. Moriyama, Y. Moruyama, Y. Yang and J.Sugimato, "Study on Subcooled water injection into molten material", Heat Transfer 6, 69 (1998).
5. S.J. Board, at al., "Detonation of Fuel Coolant Explosions", Nature, 254, 319 (1975).
6. B.E. Gelfand, "Droplet breakup phenomena in flows with velocity lag", Prog. Energy Combust. Sci., 22, 201 (1996).
7. S.Tomotika, "On the instability of a cylindrical thread of viscous liquid surrounded by another" Proc. Ray. Soc.(London), A150, 322 (1935).
8. B.J. Meister and G. F. Shecle, "Generalized solution of the Tomotika stability analysis for a cylindrical jet", AICRE Journal, 13, 4 (1967).
9. A.A Shutov, "Instability of a compounded jet of drip fluids", Mechanics of fluid and gas, 4, 3 (1985).
10. A. Jeffrey, *Handbook of Mathematical Formulas and Integrals*, 2nd Edition (Newcastle Upon Tyne, VK, 2000).